\documentclass[%
 prl ,
 amsmath,amssymb,
 reprint,%
]{revtex4-2}

\usepackage{figsize}

\usepackage{amsbsy}
\usepackage{graphicx}
\usepackage{dcolumn}
\usepackage{bm}
\usepackage{amsmath,amsthm,amssymb}
\usepackage{tikz}
\usepackage{braket}
\usepackage{comment}
\usepackage{mathrsfs}
\usepackage{stackengine}

\newcommand{\bD}{\mathbf{D}}
\newcommand{\bE}{\mathbf{E}}

\newcommand{\be}{\mathbf{e}}

\newcommand{\bh}{\mathbf{h}}
\newcommand{\bj}{\mathbf{j}}

\newcommand{\br}{\mathbf{r}}

\newcommand{\cE}{\mathcal{E}}

\newcommand{\bcG}{\bar{\mathcal{G}}}
\newcommand{\bbcG}{\bar{\boldsymbol{\mathcal{G}}}}

\newcommand{\cH}{\mathcal{H}}
\newcommand{\cJ}{\mathcal{J}}

\newcommand{\bbcL}{\boldsymbol{\bar{\mathcal{L}}}}

\newcommand{\bcH}{\boldsymbol{\mathcal{H}}}

\newcommand{\bcE}{\boldsymbol{\mathcal{E}}}
\newcommand{\bcJ}{\boldsymbol{\mathcal{J}}}

\newcommand{\hbphi}{\hat{\boldsymbol{\phi}}}

\newcommand{\hbz}{\hat{\mathbf{z}}}

\newcommand{\bbF}{\bar{\mathbf{F}}}

\newcommand{\bbK}{\bar{\mathbf{K}}}

\DeclareMathOperator*{\argmax}{arg\,max} 
\def\subrangle#1{\stackengine{4pt}{}{$\!\scriptstyle #1$}{U}{l}{F}{F}{L}}

\newcommand{\vbD}{\overleftarrow{\bD}}

\begin{document}

\title{Ultimate Performance of Biomedical Ablation}

\author{Sangbin Lee$^{1,2}$}
\author{Jongheon Lee$^3$}%
\author{Ada S. Y. Poon$^{4}$}
\author{Sanghoek Kim$^{1,2}$}%
\email{sanghoek@khu.ac.kr}
\affiliation{$^1$Department of Electronics and Information Convergence Engineering, Kyung Hee University, Yongin 17104, Republic of Korea\\
$^2$Institute for Wearable Convergence Electronics, Kyung Hee University, Yongin 17104, Republic of Korea\\
$^3$Ming Hsieh Department of Electrical Engineering, University of Southern California, CA 90089, USA\\
$^4$Department of Electrical Engineering, Stanford University, CA 94305, USA\\
}

\date{\today}

\begin{abstract}
Microwave ablation is a therapeutic procedure to eliminate abnormal tissue within a body selectively. There are two types of ablations; the thermal one aims to raise the temperature at the target, while the non-thermal one induces a temporarily high electric field at the target to disrupt cellular membrane integrity. This work identifies the fundamental bounds of the efficiency for each type of ablation and the sources to achieve them. 
For both types, the bounds exceed the performance of existing solutions by tenfold, showing a large room for improvement. 
Finally, the optimal source for thermal ablation is physically realized with an $11\times11$ dipole array, the performance of which closely approaches the bound.

\end{abstract}

\keywords{Hyperthermia, SAR, Bioheat Equation}
\maketitle

\textit{Introduction.}\---- Focusing the field in a confined region has been a long-sought-after goal in both microwave~\cite{pfeiffer2013metamaterial, grbic2004overcoming} and optical~\cite{pendry2000negative, fang2005sub} spectrums due to its various applications, such as high-resolution microscopes~\cite{fang2005sub}, lithography~\cite{timp1992using},  biomedicines~\cite{stang2012preclinical, yang2020antireflection, ho2015planar}, as well as wireless power transfer systems~\cite{kim2012wireless, kim2013midfield}.  Among those,  microwave ablation for biomedical applications is different from the others; (\textit{i}) the human tissue in which the wave travels is a lossy medium, (\textit{ii}) the depth of target can be comparable to the source size, which is limited by the body scale. These distinctive features make focusing methods for other applications not directly applicable to the biomedical ablation. 

Biomedical ablations can be categorized into two types according to their purpose. The first type is the \textit{thermal ablation}, also known as hyperthermia, which concentrates field \textit{spatially} at a target area for nearly an hour to gradually elevate temperature of the target up to 40 $\sim$ 43~$^\circ$C, so that apoptotic and necrotic cell death may occur in the target area~\cite{hildebrandt2002cellular} or make the cells more vulnerable to radio~\cite{group1996radiotherapy} or chemotherapy~\cite{vujaskovic2010phase}.
The other type is a \textit{non-thermal ablation}, of which the purpose is to induce high electric field momentarily at a target leading to a sudden disruption of the cell homeostasis~\cite{verma2021primer}, for which the \textit{spatiotemporal} focusing of the field is necessary.

\begin{figure}
	\centering
	\includegraphics[width=0.5\textwidth]{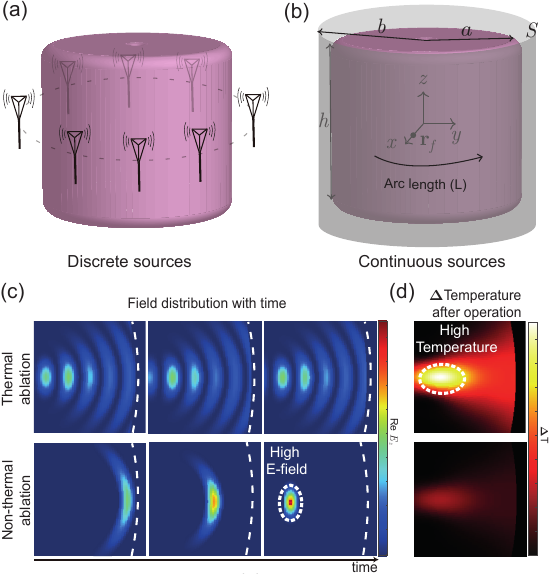} 
	\caption{(a) Discrete sources, (b) continuously distributed sources on a sheet. (c) Fields according to time and (d) temperature variations of the thermal (top) and the non-thermal (bottom) ablations.}
	\label{fig:configuration}
\end{figure}

Over decades, various kinds of ablation applicators have been developed. 
Although the applicator with a single source could successfully destroy a superficial target, it is incapable of localizing the heat at deep tissue~\cite{stauffer2000thermal, wust2002hyperthermia}.
In an effort to deepen the target point, the coherent array of sources placed around the surface of tissue was proposed (Fig.~\ref{fig:configuration}(a)). 
Refs.~\cite{gee1984focused, ling1984focusing} optimized the magnitude and phase of each source using the time-reversal method, in which the electric field at the target from each source should interfere in-phase. For ablation, however, it is more directly associated with its purpose to optimize the \textit{focusing gain}, which is defined as the ratio of the electric-field strength at the target to the average electric strength across the body. 
For example, refs.~\cite{seebass2001electromagnetic, keith1999optimization, lee2023discrete} deduced the source excitation to maximize the gain of time-averaged strength for the thermal ablation.

In previous works, however, the sources were restricted as the array of discrete elements (Fig.~\ref{fig:configuration}(a)) with a specific type, such as horn~\cite{gee1984focused}, dipole antenna~\cite{seebass2001electromagnetic, keith1999optimization, lee2023discrete}, or electric current filaments~\cite{boag1990optimal, boag1993analysis}, which prevents the results from being the global optimum. Since any physical antenna has an inherent dependency on frequency, the optimal frequency obtained over a confined source domain cannot be claimed as the global optimum, either~\cite{ling1984frequency, lee2023discrete, seebass2001electromagnetic, keith1999optimization}. As a result, currently, various types of sources are employed in practice with their operating frequency lying from a low megahertz to a gigahertz range~\cite{wust2002hyperthermia, vander2006rf} in the lack of convincing arguments that which one should yield the global optimum.

In contrast, this Letter finds the solution that maximizes the focusing gain from the general source configuration shown in Fig.~\ref{fig:configuration}(b). 
The source is modeled by the continuous electric current density on the surface $S$ surrounding the cylindrical tissue, without any other constraint.
By the field equivalence principle, such a surface current source can represent an arbitrary external source in the aspect of field distribution inside the tissue. Therefore, the resultant focusing gain from the optimization of this work can indeed be the upper bound of the gain. 

For the ultimate \textit{thermal} ablation, we show the optimal source should be the monochromatic source with its frequency in a low-gigahertz range. While it lacks a temporal focusing (Fig. 1(c), top), the temperature of a target of which the depth is less than 4~cm can be efficiently elevated (Fig.~\ref{fig:configuration}(d), top) after the operation.
The global bound of the focusing gain reveals significant room for improvement over the existing solutions by 9.1~dB. 
 The physical implementation is realized by a dipole array that mimics the optimal current distribution. We demonstrate that the focusing gain of the dipole array approaches very close to its upper limit. 

 For the ultimate \textit{non-thermal} ablation, we find a transient current source to induce the spatiotemporal focusing of the electric field at a target (Fig.~\ref{fig:configuration}(c), bottom). The electric field can be focused in a few millimeter and sub-nanosecond scales, surpassing the performance of a primitive dipole-array source by 13.3~dB in its focusing gain. However, because the duration of high electric field is very short, the temperature rise after the operation is not considerable (Fig.~\ref{fig:configuration}(d), bottom).




\textit{Field Expression.}\---- This work models the human body as a cylindrical trunk. 
The geometric configuration is illustrated in Fig.~\ref{fig:configuration}(b), where the radius and the height of the trunk are $a$ and $h$, respectively. 
To simplify the analysis, the height $h$ is assumed to be infinite in the field computation, and the dielectric properties of the trunk are considered to be a homogeneous muscle with a complex permittivity $\epsilon_1$ and permeability $\mu$.

We aim to find a source outside the body to focus the electric field at a focal point $\br_f$. By the equivalence principle, one can always find a tangential electrical current density on a coaxial cylindrical surface with the radius $b\;(>a)$ surrounding the body trunk, which creates identical fields with such an optimal one. Therefore, without loss of generality, we may assume the surface-current source $\bj(\phi, z,t)$ is confined to the cylindrical surface with the radius $b$; $\bj(\phi, z, t) = j_\phi(\phi, z, t)\hbphi + j_z(\phi, z,t)\hbz$. The current source creates the electric field $\be(\br, t)$ and displacement current $\bj_D(\br, t)$ within the tissue.

Within the trunk, the fields can be decomposed into elementary wave functions $\psi_{n, k_z, \omega}(\phi, z, t) =e^{-jn\phi-jk_z z+j\omega t}$~\cite{chew1995waves, harrington2001time};
\begin{align}
\begin{bmatrix}
\be(\br, t) \\
\bh(\br, t)
\end{bmatrix} 
= \frac{1}{2\pi} \sum_{n}\iint dk_z d\omega 
\begin{bmatrix}
\bcE(\rho, n, k_z, \omega) \psi_{n, k_z, \omega}\\
\bcH(\rho, n, k_z, \omega) \psi_{n, k_z, \omega}
\end{bmatrix},
\end{align}
where $\rho=\sqrt{x^2+y^2}$, $\omega$ is the radial frequency,  $n$ is the order of the Bessel's equation, and $k_z$ is the wave number in the $\hat{z}$ direction; $k^2_{1\rho}+k^2_z=\omega^2\mu\epsilon_1$. 
The set of $\{ \psi_{n, k_z, \omega} \}$ is the orthogonal basis for the Hilbert space $\Psi$ forming the entire Helmholtz equation solutions for the cylindrical surface with the radius $\rho$.

The $\bcE$ and $\bcH$ are the spectral representation of the fields that can be computed through the Green's operator $\bbcG_E$ and $\bbcG_H$ in the spectral domain~\cite{chew1995waves} which maps the space of source $\Omega = \{\bcJ | \bcJ=[\cJ_\phi, \cJ_z]^T\}$ into $\Psi$; for example, 
\begin{align}
	\bcE =  \bbcG_E\bcJ 
	= - \frac{\pi b}{2 \omega \epsilon_1} \bbK \left(\bbF_n \vbD_s'\right)\bcJ.
	\label{EvsJ}
\end{align} 
In \eqref{EvsJ},  the linear operator $\bbK$  associates the $\hat{z}$ component of the fields $[\cE_z, \cH_z]$ with $\bcE$, $\bbF_n$ accounts for the transmission coefficient of the $n$-th mode across the interface, and $\vbD_s'$ is a differential operator acting on functions to its left to express the emanation of the fields by the currents. It should be noted that the $\bbcG_E$ is a diagonal operator, which implies the $(n, k_z, \omega)$-mode of the current $\bcJ$ emanates only the corresponding $(n, k_z, \omega)$-mode of the fields. 
The expressions for the operators are given in the Supplemental Materials I.

\begin{figure*}[t]
\vspace{-20pt}
	\centering
	\includegraphics[width=0.9\textwidth]{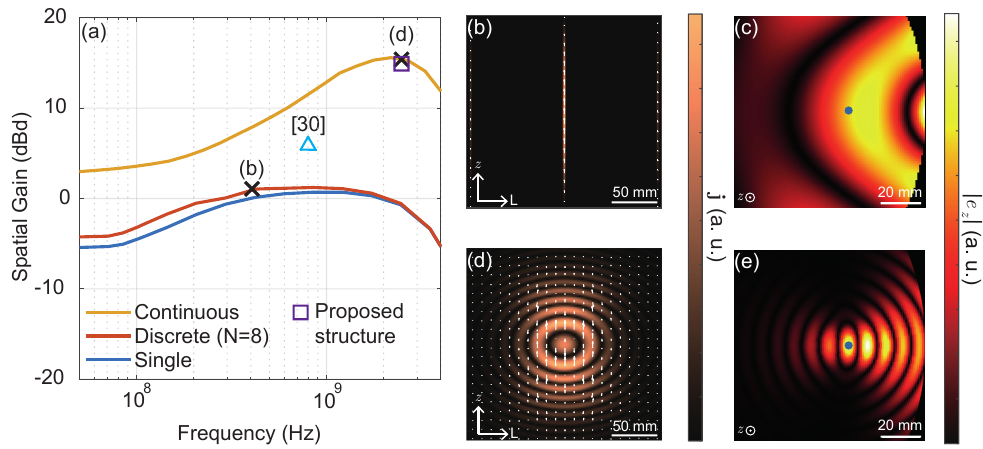}
	\caption{(a) The spatial gains of various source types according to the operating frequency. The sky-blue triangle and the purple rectangle indicate the gain of~\cite{song2022micro} and the proposed physical structure in this work, respectively. (b) The current distribution on the cylindrical surface $S$ of an eight equally-spaced dipole array~\cite{lee2023discrete} at 400~MHz and (c) its electric-field strength in the body. (d) The current distribution of the spatially optimal source on the cylindrical surface $S$ at 2.45~GHz and (e) its electric-field strength in the body. The blue dots in (c) and (e) indicate the position of the focal point. }
	\label{Figure2}
\end{figure*}

\textit{Figures of merit.}\----Two figures of merit are defined to describe the performance of the ablation. One is the spatial (focusing) gain $F_{s}$ indicating the performance of the thermal ablation, which is the ratio of the time-averaged thermal loss at target $\br_f$ to the average thermal loss in the body of the volume $V$; 
\begin{align}
	F_{s}&= \frac{\int dt \; \be(\br_f, t) \cdot\bj_D(\br_f, t)}{\iint d\br dt \; \be(\br, t)\cdot\bj_D(\br, t)/V}\label{spatial gain}\\
	&= \frac{\int d\omega \; \sigma(\omega) |\bE(\br_f, \omega)|^2}{\iint d\br d\omega \; \sigma(\omega)|\bE(\br, \omega)|^2/V}, \nonumber
\end{align} 
where $\sigma$ is the conductivity of a tissue~\cite{gabriel1996dielectric} and $\bE(\br,\omega)$ is the Fourier transform of the electric field $\be(\br, t)$. 
As we shall see, one can deduce the frequency $\omega_{opt}$ that maximizes the focusing gain: $\omega_{opt} = \argmax_{\omega} \frac{V|\bE(\mathbf{r}_f, \omega)|^2}{\int_V d\br\; |\bE(\mathbf{r},\omega)|^2}$. This makes the spatial gain to be bounded by the gain of the monochromatic field at $\omega_{opt}$. 
Moreover, the characteristic of cylindrical structure makes the electric field can be the strongest along the $\hbz$ direction (see the Supplemental Materials II.B and \cite{iero2015role}), leading the spatial gain to be bounded by
\begin{align}
	F_{s} \leq  \frac{V |E_z(\mathbf{r}_f, \omega_{opt}) |^2}{\int_V d\br\;  |\bE(\mathbf{r},\omega_{opt})|^2}.
	\label{spatial gain bound}
\end{align} 

The second figure of merit is the spatiotemporal (focusing) gain $F_{st}$ representing the performance of the non-thermal ablation. It is the ratio of the instantaneous electric-field strength to the overall electric strength across the body and time; 
 \begin{align}
	F_{st} &=  \frac{V|e_z(\br_f, t_f=0)|^2}{\iint d\br dt \; |\be(\br, t)|^2}  = \frac{V}{2\pi}\frac{|\int d\omega \;  E_z(\br_f, \omega)|^2}{\iint d\br d\omega \; |\bE(\br, \omega)|^2},
	\label{spatiotemporal gain}
\end{align} 
where we assume the electric field at target has a peak at time $t_f=0$ without the loss of generality. Likewise, our coordinate system is aligned to make $\phi_{f}=z_{f}=0$ of $\br_f$ to simplify the expression. 





\textit{Source Optimization}\---- 
We aim to find the upper bound of both $F_{s}$ and $F_{st}$.
By applying Parseval's theorem, both \eqref{spatial gain bound} and \eqref{spatiotemporal gain} can be expressed as the ratio of inner products in the following form (see the Supplemental Materials II.A); 
\begin{equation}
F=\frac{V}{4\pi^2}\frac{\left|\langle 1, \bcG_{E_z}\bcJ\rangle \right|^2}{\langle \bcJ,  \bbcL\bbcG^\dagger_{E}\bbcG_{E}\bcJ\rangle},
\label{form}
\end{equation} 
where $\bbcL = \int_0^a \rho d\rho$ is the integral operator along the radial direction. Although the appearances are identical, the difference between the two gains lies in the definition of the inner products. For the spatial gain with a monochromatic field at $\omega_0$, the inner product is defined as $\langle \bcJ_1, \bcJ_2 \rangle\subrangle{\Omega_0} :=  \sum_n \int dk_z  \left[ \cJ_{1, \phi}^*  \cJ_{2, \phi} + \cJ_{1, z}^*  \cJ_{2, z} \right]$, where the space for the inner product is the subspace $\Omega_0$ ($\subseteq \Omega$)  in which the radial frequency is fixed as $\omega_0$. On the other hand, for the spatiotemporal gain, the space for the inner product expands as the entire space of source $\Omega$ and the inner product is defined as $\langle \bcJ_1, \bcJ_2 \rangle\subrangle{\Omega} := \frac{1}{2\pi} \sum_n \int dk_z \int d\omega  \left[ \cJ_{1, \phi}^*  \cJ_{2, \phi} + \cJ_{1, z}^*  \cJ_{2, z} \right]$. No matter how the inner product is defined, the bound of \eqref{form} can be analytically computed by the Cauchy-Schwartz inequality as 
\begin{equation}
F_{opt} = \frac{V}{4\pi^2}\langle  \bcG^\dagger_{E_z},  (\bbcL \bbcG^\dagger_E \bbcG_E)^{-1}\bcG^\dagger_{E_z}\rangle
\label{bound}
\end{equation}
with the corresponding definition of the inner product.

\begin{figure*}[t]
\vspace{-20pt}
	\centering
	\includegraphics[width=0.9\textwidth]{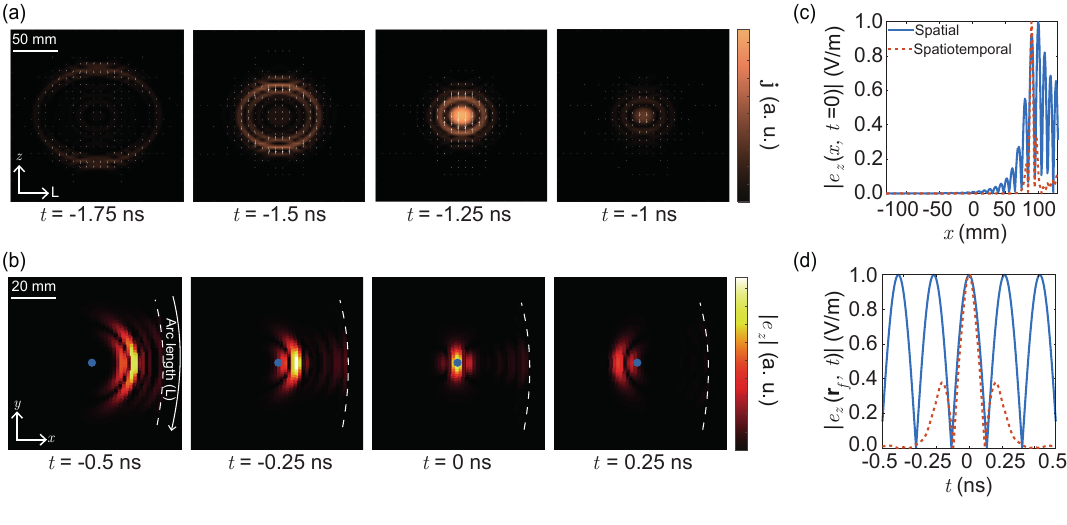}
	\caption{(a) Current distribution of the spatiotemporally optimal source and (b) its instantaneous electric field in the body over time. The blue dots indicate the focal point $\br_f$. (c) Instantaneous electric-field strength along the $x$-axis at time $t=t_f$($=0$~ns)  and (d) along the time at the focal point $\br_f$ for the spatially and the spatiotemporally optimal sources.}
	\label{Figure3}
\end{figure*}

\textit{Numerical Results.}\---- For numerical demonstration, the parameters are set to $(\rho_f, a, b)=(9, 13, 14)$~cm. 

The spatial gain is normalized by that of a single $\lambda/2$-dipole source at 400~MHz. The gain has the unit of ``dBd" to denote that it is a relative gain compared to the dipole source.
The spatial gains for various configurations of the source are shown in Fig.~\ref{Figure2}(a). For 4($=a-\rho_f$)-cm focal depth, the eight-dipole source in Fig.~\ref{Figure2}(b) of which the excitations are optimized~\cite{lee2023discrete} barely improves the spatial gain (the cross mark on the red curve in Fig.~\ref{Figure2}(a)). Fig.~\ref{Figure2}(a) also presents the spatial gain of a highly directive waveguide antenna for hyperthermia at 915~MHz~\cite{song2022micro}. Compared to them, the optimal continuous source shown in Fig.~\ref{Figure2}(d) demonstrates the room for the gain improvement by 14.4~dB and 9.1~dB, respectively. 

The optimal operating frequency $\omega_{opt}$ lies in a low-GHz range, at which the short wavelength greatly enhances the spatial gain. Such a focusing effect obviously appears by comparing the electric-field strength by the eight-dipole source at 400~MHz and that by the optimal source shown in Fig.~\ref{Figure2}(c) and (e), respectively.
Supplemental Fig.~S4 shows that the optimal frequency remains in a low-GHz range for various focal depths varying from 2~cm to 12~cm. 
Beyond the low-gigahertz range, the conductivity of tissue abruptly increases and deteriorates the spatial gain~\cite{meng2010optimal}. 

The maximum depth for hyperthermia can be also investigated by solving the Penne's bioheat equation. It shows that up to the focal depth of 4~cm, the temperature can peak at the desired depth after an hour's operation with the ultimate thermal ablation (See the Supplemental Materials IV).

We numerically demonstrate the bound of spatiotemporal gain under the same configuration of $(\rho_f, a, b)$ as the spatial one. Fig.~\ref{Figure3}(a) shows the optimal source should flow inward radially, which makes the electric field spatially focused at a region with the size 5$\times$10$\times$10~mm$^3$ at $t=0$~ns, visualized in Fig.~\ref{Figure3}(b) and (c). 
In contrast to the field by the spatially optimal source, the field fades out after having the peak for a few tenths of a nanosecond (Fig.~\ref{Figure3}(d)) to minimize the total loss in tissue across the time. 
The sub-nanosecond duration of the peak indicates that the spatiotemporally optimal field is also dominated by low-GHz components like the spatially optimal field. 
A higher frequency component than that does not help to enhance the gain $F_{st}$ because it severely attenuates before it reaches the target.  
Quantitatively, the spatiotemporal gain of the optimal source is 13.3~dB higher than that of the eight-dipole array (See the Supplemental Materials~V).
\begin{figure}[!h]
	\centering
	\includegraphics[width=0.5\textwidth]{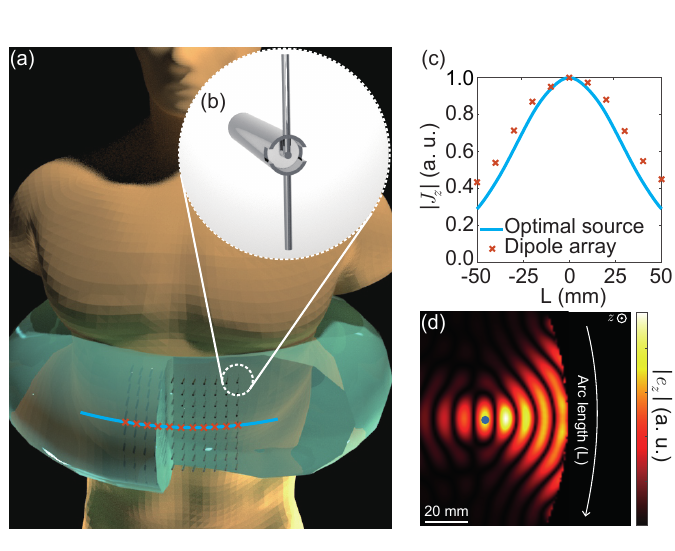}
	\caption{(a) Physical implementation of the source as an 11$\times$11 balanced, half-wavelength dipole array of (b), immersed in the water bolus. (c) The magnitude of excitation coefficients (red cross marks) of the dipole array and the spatially optimal source (blue curve) along the arc length $L$ at $z=0$. (d) Electric-field strength emanated by the proposed 11$\times$11 dipole array. The blue dot indicates the focal point $\br_f$. }
	\label{Figure4}
\end{figure}

\vspace{3pt}
\textit{Physical Implementation.}\---- This section investigates the physical implementation of the optimal current distribution for thermal ablation. We demonstrate the developmental procedure of the physical antenna that mimics the optimal distribution targeting the depth of 4~cm. As observed in Fig.~\ref{Figure2}(d), the $\hat{z}$ component dominates in the optimal current distribution. 
 To synthesize it, we adopt an 11-by-11 phased-array antenna over 10$\times$10~cm$^2$ area, depicted in Fig.~\ref{Figure4}(a), where each element is the balanced, half-wavelength dipole antenna (6~mm length at 2.45~GHz in water, Fig.~\ref{Figure4}(b)) along the $\hat{z}$ direction. The spacing between the elements is set to be 10~mm.

The excitation coefficients of the dipole array are determined as follows. 
From the principle of superposition, the net electric field can be expressed as the excitation coefficients multiplied by a field matrix. The component of the field matrix describes the electric field distribution of each dipole and can be obtained by a commercial electromagnetic simulator, the CST Microwave. 
Based on the field matrix, the excitation coefficients of the array can be numerically decided to maximize the field strength at the focal point for a given total loss in the tissue~\cite{kim2012wireless2, seebass2001electromagnetic, keith1999optimization}. The Fig.~\ref{Figure4}(c) shows the coefficients obtained as such along $z=0$ (the red crosses), closely following the optimal solution over the continuous domain (the blue curve). The field distribution (Fig.~\ref{Figure4}(d)) inside the body by the dipole array also discloses the focusing effect approximating the optimal one in Fig.~\ref{Figure2}(e), achieving its focusing gain (the purple-square mark in Fig.~\ref{Figure2}(a)) only 0.4~dB below the upper bound. As marked in Fig.~\ref{Figure2}(a), this is a superior performance to any previous work by far.

\textit{Conclusion.}\---- Previously, the domains of the optimization for hyperthermia were inherently restricted, failing to claim the global optimum. This work performs the optimization over a hypothetical, continuous current density, deducing the globally optimal current distribution, and addressing the ultimate performance for both thermal and non-thermal ablations. 

As a numerical example, the performance of the ablations is demonstrated when the focal depth is 4~cm.  The bound of the thermal gain reveals the room for improvement over the existing solutions by 9.1~dB.  Likewise, the bound of the non-thermal gain exceeds the gain of a primitive source, the equally-spaced eight-dipole array, by 13.3~dB.
\section{acknowledgments}
\begin{acknowledgments}
This work was in part supported by the Ministry of Science, ICT and
Future Planning, Korea, under the Information Technology Research Center support program (No. IITP-2021-0-02046)
supervised by the IITP (National IT Industry Promotion Agency), in part supported by the National Research Foundation of Korea (No.  NRF2018R1A6A1A03025708 and NRF-2023R1A2C2004236).
\end{acknowledgments}

\appendix


%
%
%


\bibliography{reference}

\begin{thebibliography}{32}%
\makeatletter
\providecommand \@ifxundefined [1]{%
 \@ifx{#1\undefined}
}%
\providecommand \@ifnum [1]{%
 \ifnum #1\expandafter \@firstoftwo
 \else \expandafter \@secondoftwo
 \fi
}%
\providecommand \@ifx [1]{%
 \ifx #1\expandafter \@firstoftwo
 \else \expandafter \@secondoftwo
 \fi
}%
\providecommand \natexlab [1]{#1}%
\providecommand \enquote  [1]{``#1''}%
\providecommand \bibnamefont  [1]{#1}%
\providecommand \bibfnamefont [1]{#1}%
\providecommand \citenamefont [1]{#1}%
\providecommand \href@noop [0]{\@secondoftwo}%
\providecommand \href [0]{\begingroup \@sanitize@url \@href}%
\providecommand \@href[1]{\@@startlink{#1}\@@href}%
\providecommand \@@href[1]{\endgroup#1\@@endlink}%
\providecommand \@sanitize@url [0]{\catcode `\\12\catcode `\$12\catcode
  `\&12\catcode `\#12\catcode `\^12\catcode `\_12\catcode `\%12\relax}%
\providecommand \@@startlink[1]{}%
\providecommand \@@endlink[0]{}%
\providecommand \url  [0]{\begingroup\@sanitize@url \@url }%
\providecommand \@url [1]{\endgroup\@href {#1}{\urlprefix }}%
\providecommand \urlprefix  [0]{URL }%
\providecommand \Eprint [0]{\href }%
\providecommand \doibase [0]{https://doi.org/}%
\providecommand \selectlanguage [0]{\@gobble}%
\providecommand \bibinfo  [0]{\@secondoftwo}%
\providecommand \bibfield  [0]{\@secondoftwo}%
\providecommand \translation [1]{[#1]}%
\providecommand \BibitemOpen [0]{}%
\providecommand \bibitemStop [0]{}%
\providecommand \bibitemNoStop [0]{.\EOS\space}%
\providecommand \EOS [0]{\spacefactor3000\relax}%
\providecommand \BibitemShut  [1]{\csname bibitem#1\endcsname}%
\let\auto@bib@innerbib\@empty
\bibitem [{\citenamefont {Pfeiffer}\ and\ \citenamefont
  {Grbic}(2013)}]{pfeiffer2013metamaterial}%
  \BibitemOpen
  \bibfield  {author} {\bibinfo {author} {\bibfnamefont {C.}~\bibnamefont
  {Pfeiffer}}\ and\ \bibinfo {author} {\bibfnamefont {A.}~\bibnamefont
  {Grbic}},\ }\bibfield  {title} {\bibinfo {title} {Metamaterial huygens’
  surfaces: tailoring wave fronts with reflectionless sheets},\ }\href@noop {}
  {\bibfield  {journal} {\bibinfo  {journal} {Physical review letters}\
  }\textbf {\bibinfo {volume} {110}},\ \bibinfo {pages} {197401} (\bibinfo
  {year} {2013})}\BibitemShut {NoStop}%
\bibitem [{\citenamefont {Grbic}\ and\ \citenamefont
  {Eleftheriades}(2004)}]{grbic2004overcoming}%
  \BibitemOpen
  \bibfield  {author} {\bibinfo {author} {\bibfnamefont {A.}~\bibnamefont
  {Grbic}}\ and\ \bibinfo {author} {\bibfnamefont {G.~V.}\ \bibnamefont
  {Eleftheriades}},\ }\bibfield  {title} {\bibinfo {title} {Overcoming the
  diffraction limit with a planar left-handed transmission-line lens},\
  }\href@noop {} {\bibfield  {journal} {\bibinfo  {journal} {Physical review
  letters}\ }\textbf {\bibinfo {volume} {92}},\ \bibinfo {pages} {117403}
  (\bibinfo {year} {2004})}\BibitemShut {NoStop}%
\bibitem [{\citenamefont {Pendry}(2000)}]{pendry2000negative}%
  \BibitemOpen
  \bibfield  {author} {\bibinfo {author} {\bibfnamefont {J.~B.}\ \bibnamefont
  {Pendry}},\ }\bibfield  {title} {\bibinfo {title} {Negative refraction makes
  a perfect lens},\ }\href@noop {} {\bibfield  {journal} {\bibinfo  {journal}
  {Physical review letters}\ }\textbf {\bibinfo {volume} {85}},\ \bibinfo
  {pages} {3966} (\bibinfo {year} {2000})}\BibitemShut {NoStop}%
\bibitem [{\citenamefont {Fang}\ \emph {et~al.}(2005)\citenamefont {Fang},
  \citenamefont {Lee}, \citenamefont {Sun},\ and\ \citenamefont
  {Zhang}}]{fang2005sub}%
  \BibitemOpen
  \bibfield  {author} {\bibinfo {author} {\bibfnamefont {N.}~\bibnamefont
  {Fang}}, \bibinfo {author} {\bibfnamefont {H.}~\bibnamefont {Lee}}, \bibinfo
  {author} {\bibfnamefont {C.}~\bibnamefont {Sun}},\ and\ \bibinfo {author}
  {\bibfnamefont {X.}~\bibnamefont {Zhang}},\ }\bibfield  {title} {\bibinfo
  {title} {Sub-diffraction-limited optical imaging with a silver superlens},\
  }\href@noop {} {\bibfield  {journal} {\bibinfo  {journal} {science}\ }\textbf
  {\bibinfo {volume} {308}},\ \bibinfo {pages} {534} (\bibinfo {year}
  {2005})}\BibitemShut {NoStop}%
\bibitem [{\citenamefont {Timp}\ \emph {et~al.}(1992)\citenamefont {Timp},
  \citenamefont {Behringer}, \citenamefont {Tennant}, \citenamefont
  {Cunningham}, \citenamefont {Prentiss},\ and\ \citenamefont
  {Berggren}}]{timp1992using}%
  \BibitemOpen
  \bibfield  {author} {\bibinfo {author} {\bibfnamefont {G.}~\bibnamefont
  {Timp}}, \bibinfo {author} {\bibfnamefont {R.}~\bibnamefont {Behringer}},
  \bibinfo {author} {\bibfnamefont {D.}~\bibnamefont {Tennant}}, \bibinfo
  {author} {\bibfnamefont {J.}~\bibnamefont {Cunningham}}, \bibinfo {author}
  {\bibfnamefont {M.}~\bibnamefont {Prentiss}},\ and\ \bibinfo {author}
  {\bibfnamefont {K.}~\bibnamefont {Berggren}},\ }\bibfield  {title} {\bibinfo
  {title} {Using light as a lens for submicron, neutral-atom lithography},\
  }\href@noop {} {\bibfield  {journal} {\bibinfo  {journal} {Physical review
  letters}\ }\textbf {\bibinfo {volume} {69}},\ \bibinfo {pages} {1636}
  (\bibinfo {year} {1992})}\BibitemShut {NoStop}%
\bibitem [{\citenamefont {Stang}\ \emph {et~al.}(2012)\citenamefont {Stang},
  \citenamefont {Haynes}, \citenamefont {Carson},\ and\ \citenamefont
  {Moghaddam}}]{stang2012preclinical}%
  \BibitemOpen
  \bibfield  {author} {\bibinfo {author} {\bibfnamefont {J.}~\bibnamefont
  {Stang}}, \bibinfo {author} {\bibfnamefont {M.}~\bibnamefont {Haynes}},
  \bibinfo {author} {\bibfnamefont {P.}~\bibnamefont {Carson}},\ and\ \bibinfo
  {author} {\bibfnamefont {M.}~\bibnamefont {Moghaddam}},\ }\bibfield  {title}
  {\bibinfo {title} {A preclinical system prototype for focused microwave
  thermal therapy of the breast},\ }\href@noop {} {\bibfield  {journal}
  {\bibinfo  {journal} {IEEE Transactions on Biomedical Engineering}\ }\textbf
  {\bibinfo {volume} {59}},\ \bibinfo {pages} {2431} (\bibinfo {year}
  {2012})}\BibitemShut {NoStop}%
\bibitem [{\citenamefont {Yang}\ \emph {et~al.}(2020)\citenamefont {Yang},
  \citenamefont {Raeker}, \citenamefont {Nguyen}, \citenamefont {Miller},
  \citenamefont {Xiong}, \citenamefont {Grbic},\ and\ \citenamefont
  {Ho}}]{yang2020antireflection}%
  \BibitemOpen
  \bibfield  {author} {\bibinfo {author} {\bibfnamefont {F.}~\bibnamefont
  {Yang}}, \bibinfo {author} {\bibfnamefont {B.~O.}\ \bibnamefont {Raeker}},
  \bibinfo {author} {\bibfnamefont {D.~T.}\ \bibnamefont {Nguyen}}, \bibinfo
  {author} {\bibfnamefont {J.~D.}\ \bibnamefont {Miller}}, \bibinfo {author}
  {\bibfnamefont {Z.}~\bibnamefont {Xiong}}, \bibinfo {author} {\bibfnamefont
  {A.}~\bibnamefont {Grbic}},\ and\ \bibinfo {author} {\bibfnamefont {J.~S.}\
  \bibnamefont {Ho}},\ }\bibfield  {title} {\bibinfo {title} {Antireflection
  and wavefront manipulation with cascaded metasurfaces},\ }\href@noop {}
  {\bibfield  {journal} {\bibinfo  {journal} {Physical Review Applied}\
  }\textbf {\bibinfo {volume} {14}},\ \bibinfo {pages} {064044} (\bibinfo
  {year} {2020})}\BibitemShut {NoStop}%
\bibitem [{\citenamefont {Ho}\ \emph {et~al.}(2015)\citenamefont {Ho},
  \citenamefont {Qiu}, \citenamefont {Tanabe}, \citenamefont {Yeh},
  \citenamefont {Fan},\ and\ \citenamefont {Poon}}]{ho2015planar}%
  \BibitemOpen
  \bibfield  {author} {\bibinfo {author} {\bibfnamefont {J.~S.}\ \bibnamefont
  {Ho}}, \bibinfo {author} {\bibfnamefont {B.}~\bibnamefont {Qiu}}, \bibinfo
  {author} {\bibfnamefont {Y.}~\bibnamefont {Tanabe}}, \bibinfo {author}
  {\bibfnamefont {A.~J.}\ \bibnamefont {Yeh}}, \bibinfo {author} {\bibfnamefont
  {S.}~\bibnamefont {Fan}},\ and\ \bibinfo {author} {\bibfnamefont {A.~S.}\
  \bibnamefont {Poon}},\ }\bibfield  {title} {\bibinfo {title} {Planar
  immersion lens with metasurfaces},\ }\href@noop {} {\bibfield  {journal}
  {\bibinfo  {journal} {Physical Review B}\ }\textbf {\bibinfo {volume} {91}},\
  \bibinfo {pages} {125145} (\bibinfo {year} {2015})}\BibitemShut {NoStop}%
\bibitem [{\citenamefont {Kim}\ \emph {et~al.}(2012{\natexlab{a}})\citenamefont
  {Kim}, \citenamefont {Ho},\ and\ \citenamefont {Poon}}]{kim2012wireless}%
  \BibitemOpen
  \bibfield  {author} {\bibinfo {author} {\bibfnamefont {S.}~\bibnamefont
  {Kim}}, \bibinfo {author} {\bibfnamefont {J.~S.}\ \bibnamefont {Ho}},\ and\
  \bibinfo {author} {\bibfnamefont {A.~S.}\ \bibnamefont {Poon}},\ }\bibfield
  {title} {\bibinfo {title} {Wireless power transfer to miniature implants:
  Transmitter optimization},\ }\href@noop {} {\bibfield  {journal} {\bibinfo
  {journal} {IEEE Transactions on Antennas and Propagation}\ }\textbf {\bibinfo
  {volume} {60}},\ \bibinfo {pages} {4838} (\bibinfo {year}
  {2012}{\natexlab{a}})}\BibitemShut {NoStop}%
\bibitem [{\citenamefont {Kim}\ \emph {et~al.}(2013)\citenamefont {Kim},
  \citenamefont {Ho},\ and\ \citenamefont {Poon}}]{kim2013midfield}%
  \BibitemOpen
  \bibfield  {author} {\bibinfo {author} {\bibfnamefont {S.}~\bibnamefont
  {Kim}}, \bibinfo {author} {\bibfnamefont {J.~S.}\ \bibnamefont {Ho}},\ and\
  \bibinfo {author} {\bibfnamefont {A.~S.}\ \bibnamefont {Poon}},\ }\bibfield
  {title} {\bibinfo {title} {Midfield wireless powering of subwavelength
  autonomous devices},\ }\href@noop {} {\bibfield  {journal} {\bibinfo
  {journal} {Physical review letters}\ }\textbf {\bibinfo {volume} {110}},\
  \bibinfo {pages} {203905} (\bibinfo {year} {2013})}\BibitemShut {NoStop}%
\bibitem [{\citenamefont {Hildebrandt}\ \emph {et~al.}(2002)\citenamefont
  {Hildebrandt}, \citenamefont {Wust}, \citenamefont {Ahlers}, \citenamefont
  {Dieing}, \citenamefont {Sreenivasa}, \citenamefont {Kerner}, \citenamefont
  {Felix},\ and\ \citenamefont {Riess}}]{hildebrandt2002cellular}%
  \BibitemOpen
  \bibfield  {author} {\bibinfo {author} {\bibfnamefont {B.}~\bibnamefont
  {Hildebrandt}}, \bibinfo {author} {\bibfnamefont {P.}~\bibnamefont {Wust}},
  \bibinfo {author} {\bibfnamefont {O.}~\bibnamefont {Ahlers}}, \bibinfo
  {author} {\bibfnamefont {A.}~\bibnamefont {Dieing}}, \bibinfo {author}
  {\bibfnamefont {G.}~\bibnamefont {Sreenivasa}}, \bibinfo {author}
  {\bibfnamefont {T.}~\bibnamefont {Kerner}}, \bibinfo {author} {\bibfnamefont
  {R.}~\bibnamefont {Felix}},\ and\ \bibinfo {author} {\bibfnamefont
  {H.}~\bibnamefont {Riess}},\ }\bibfield  {title} {\bibinfo {title} {The
  cellular and molecular basis of hyperthermia},\ }\href@noop {} {\bibfield
  {journal} {\bibinfo  {journal} {Critical reviews in oncology/hematology}\
  }\textbf {\bibinfo {volume} {43}},\ \bibinfo {pages} {33} (\bibinfo {year}
  {2002})}\BibitemShut {NoStop}%
\bibitem [{\citenamefont {Group}\ \emph {et~al.}(1996)\citenamefont {Group},
  \citenamefont {Vernon}, \citenamefont {Hand}, \citenamefont {Field},
  \citenamefont {Machin}, \citenamefont {Whaley}, \citenamefont {van~der Zee},
  \citenamefont {van Putten}, \citenamefont {van Rhoon}, \citenamefont {van
  Dijk} \emph {et~al.}}]{group1996radiotherapy}%
  \BibitemOpen
  \bibfield  {author} {\bibinfo {author} {\bibfnamefont {I.~C.~H.}\
  \bibnamefont {Group}}, \bibinfo {author} {\bibfnamefont {C.~C.}\ \bibnamefont
  {Vernon}}, \bibinfo {author} {\bibfnamefont {J.~W.}\ \bibnamefont {Hand}},
  \bibinfo {author} {\bibfnamefont {S.~B.}\ \bibnamefont {Field}}, \bibinfo
  {author} {\bibfnamefont {D.}~\bibnamefont {Machin}}, \bibinfo {author}
  {\bibfnamefont {J.~B.}\ \bibnamefont {Whaley}}, \bibinfo {author}
  {\bibfnamefont {J.}~\bibnamefont {van~der Zee}}, \bibinfo {author}
  {\bibfnamefont {W.~L.}\ \bibnamefont {van Putten}}, \bibinfo {author}
  {\bibfnamefont {G.~C.}\ \bibnamefont {van Rhoon}}, \bibinfo {author}
  {\bibfnamefont {J.~D.}\ \bibnamefont {van Dijk}}, \emph {et~al.},\ }\bibfield
   {title} {\bibinfo {title} {Radiotherapy with or without hyperthermia in the
  treatment of superficial localized breast cancer: results from five
  randomized controlled trials},\ }\href@noop {} {\bibfield  {journal}
  {\bibinfo  {journal} {International Journal of Radiation Oncology* Biology*
  Physics}\ }\textbf {\bibinfo {volume} {35}},\ \bibinfo {pages} {731}
  (\bibinfo {year} {1996})}\BibitemShut {NoStop}%
\bibitem [{\citenamefont {Vujaskovic}\ \emph {et~al.}(2010)\citenamefont
  {Vujaskovic}, \citenamefont {Kim}, \citenamefont {Jones}, \citenamefont
  {Lan}, \citenamefont {McCall}, \citenamefont {Dewhirst}, \citenamefont
  {Craciunescu}, \citenamefont {Stauffer}, \citenamefont {Liotcheva},
  \citenamefont {Betof} \emph {et~al.}}]{vujaskovic2010phase}%
  \BibitemOpen
  \bibfield  {author} {\bibinfo {author} {\bibfnamefont {Z.}~\bibnamefont
  {Vujaskovic}}, \bibinfo {author} {\bibfnamefont {D.~W.}\ \bibnamefont {Kim}},
  \bibinfo {author} {\bibfnamefont {E.}~\bibnamefont {Jones}}, \bibinfo
  {author} {\bibfnamefont {L.}~\bibnamefont {Lan}}, \bibinfo {author}
  {\bibfnamefont {L.}~\bibnamefont {McCall}}, \bibinfo {author} {\bibfnamefont
  {M.~W.}\ \bibnamefont {Dewhirst}}, \bibinfo {author} {\bibfnamefont
  {O.}~\bibnamefont {Craciunescu}}, \bibinfo {author} {\bibfnamefont
  {P.}~\bibnamefont {Stauffer}}, \bibinfo {author} {\bibfnamefont
  {V.}~\bibnamefont {Liotcheva}}, \bibinfo {author} {\bibfnamefont
  {A.}~\bibnamefont {Betof}}, \emph {et~al.},\ }\bibfield  {title} {\bibinfo
  {title} {A phase i/ii study of neoadjuvant liposomal doxorubicin, paclitaxel,
  and hyperthermia in locally advanced breast cancer},\ }\href@noop {}
  {\bibfield  {journal} {\bibinfo  {journal} {International Journal of
  Hyperthermia}\ }\textbf {\bibinfo {volume} {26}},\ \bibinfo {pages} {514}
  (\bibinfo {year} {2010})}\BibitemShut {NoStop}%
\bibitem [{\citenamefont {Verma}\ \emph {et~al.}(2021)\citenamefont {Verma},
  \citenamefont {Asivatham}, \citenamefont {Deneke}, \citenamefont
  {Castellvi},\ and\ \citenamefont {Neal}}]{verma2021primer}%
  \BibitemOpen
  \bibfield  {author} {\bibinfo {author} {\bibfnamefont {A.}~\bibnamefont
  {Verma}}, \bibinfo {author} {\bibfnamefont {S.~J.}\ \bibnamefont
  {Asivatham}}, \bibinfo {author} {\bibfnamefont {T.}~\bibnamefont {Deneke}},
  \bibinfo {author} {\bibfnamefont {Q.}~\bibnamefont {Castellvi}},\ and\
  \bibinfo {author} {\bibfnamefont {R.~E.}\ \bibnamefont {Neal}},\ }\bibfield
  {title} {\bibinfo {title} {Primer on pulsed electrical field ablation:
  understanding the benefits and limitations},\ }\href@noop {} {\bibfield
  {journal} {\bibinfo  {journal} {Circulation: Arrhythmia and
  Electrophysiology}\ }\textbf {\bibinfo {volume} {14}},\ \bibinfo {pages}
  {e010086} (\bibinfo {year} {2021})}\BibitemShut {NoStop}%
\bibitem [{\citenamefont {Stauffer}(2000)}]{stauffer2000thermal}%
  \BibitemOpen
  \bibfield  {author} {\bibinfo {author} {\bibfnamefont {P.~R.}\ \bibnamefont
  {Stauffer}},\ }\bibfield  {title} {\bibinfo {title} {Thermal therapy
  techniques for skin and superficial tissue disease},\ }in\ \href@noop {}
  {\emph {\bibinfo {booktitle} {Matching the energy source to the clinical
  need: a critical review}}},\ Vol.\ \bibinfo {volume} {10297}\ (\bibinfo
  {organization} {SPIE},\ \bibinfo {year} {2000})\ pp.\ \bibinfo {pages}
  {321--361}\BibitemShut {NoStop}%
\bibitem [{\citenamefont {Wust}\ \emph {et~al.}(2002)\citenamefont {Wust},
  \citenamefont {Hildebrandt}, \citenamefont {Sreenivasa}, \citenamefont {Rau},
  \citenamefont {Gellermann}, \citenamefont {Riess}, \citenamefont {Felix},\
  and\ \citenamefont {Schlag}}]{wust2002hyperthermia}%
  \BibitemOpen
  \bibfield  {author} {\bibinfo {author} {\bibfnamefont {P.}~\bibnamefont
  {Wust}}, \bibinfo {author} {\bibfnamefont {B.}~\bibnamefont {Hildebrandt}},
  \bibinfo {author} {\bibfnamefont {G.}~\bibnamefont {Sreenivasa}}, \bibinfo
  {author} {\bibfnamefont {B.}~\bibnamefont {Rau}}, \bibinfo {author}
  {\bibfnamefont {J.}~\bibnamefont {Gellermann}}, \bibinfo {author}
  {\bibfnamefont {H.}~\bibnamefont {Riess}}, \bibinfo {author} {\bibfnamefont
  {R.}~\bibnamefont {Felix}},\ and\ \bibinfo {author} {\bibfnamefont
  {P.}~\bibnamefont {Schlag}},\ }\bibfield  {title} {\bibinfo {title}
  {Hyperthermia in combined treatment of cancer},\ }\href@noop {} {\bibfield
  {journal} {\bibinfo  {journal} {The lancet oncology}\ }\textbf {\bibinfo
  {volume} {3}},\ \bibinfo {pages} {487} (\bibinfo {year} {2002})}\BibitemShut
  {NoStop}%
\bibitem [{\citenamefont {Gee}\ \emph {et~al.}(1984)\citenamefont {Gee},
  \citenamefont {Lee}, \citenamefont {Bong}, \citenamefont {Cain},
  \citenamefont {Mittra},\ and\ \citenamefont {Magin}}]{gee1984focused}%
  \BibitemOpen
  \bibfield  {author} {\bibinfo {author} {\bibfnamefont {W.}~\bibnamefont
  {Gee}}, \bibinfo {author} {\bibfnamefont {S.-W.}\ \bibnamefont {Lee}},
  \bibinfo {author} {\bibfnamefont {N.~K.}\ \bibnamefont {Bong}}, \bibinfo
  {author} {\bibfnamefont {C.~A.}\ \bibnamefont {Cain}}, \bibinfo {author}
  {\bibfnamefont {R.}~\bibnamefont {Mittra}},\ and\ \bibinfo {author}
  {\bibfnamefont {R.~L.}\ \bibnamefont {Magin}},\ }\bibfield  {title} {\bibinfo
  {title} {Focused array hyperthermia applicator: Theory and experiment},\
  }\href@noop {} {\bibfield  {journal} {\bibinfo  {journal} {IEEE transactions
  on biomedical engineering}\ ,\ \bibinfo {pages} {38}} (\bibinfo {year}
  {1984})}\BibitemShut {NoStop}%
\bibitem [{\citenamefont {Ling}\ and\ \citenamefont
  {Lee}(1984)}]{ling1984focusing}%
  \BibitemOpen
  \bibfield  {author} {\bibinfo {author} {\bibfnamefont {H.}~\bibnamefont
  {Ling}}\ and\ \bibinfo {author} {\bibfnamefont {S.-W.}\ \bibnamefont {Lee}},\
  }\bibfield  {title} {\bibinfo {title} {Focusing of electromagnetic waves
  through a dielectric interface},\ }\href@noop {} {\bibfield  {journal}
  {\bibinfo  {journal} {JOSA A}\ }\textbf {\bibinfo {volume} {1}},\ \bibinfo
  {pages} {965} (\bibinfo {year} {1984})}\BibitemShut {NoStop}%
\bibitem [{\citenamefont {Seebass}\ \emph {et~al.}(2001)\citenamefont
  {Seebass}, \citenamefont {Beck}, \citenamefont {Gellermann}, \citenamefont
  {Nadobny},\ and\ \citenamefont {Wust}}]{seebass2001electromagnetic}%
  \BibitemOpen
  \bibfield  {author} {\bibinfo {author} {\bibfnamefont {M.}~\bibnamefont
  {Seebass}}, \bibinfo {author} {\bibfnamefont {R.}~\bibnamefont {Beck}},
  \bibinfo {author} {\bibfnamefont {J.}~\bibnamefont {Gellermann}}, \bibinfo
  {author} {\bibfnamefont {J.}~\bibnamefont {Nadobny}},\ and\ \bibinfo {author}
  {\bibfnamefont {P.}~\bibnamefont {Wust}},\ }\bibfield  {title} {\bibinfo
  {title} {Electromagnetic phased arrays for regional hyperthermia: optimal
  frequency and antenna arrangement},\ }\href@noop {} {\bibfield  {journal}
  {\bibinfo  {journal} {International Journal of Hyperthermia}\ }\textbf
  {\bibinfo {volume} {17}},\ \bibinfo {pages} {321} (\bibinfo {year}
  {2001})}\BibitemShut {NoStop}%
\bibitem [{\citenamefont {Keith}\ \emph {et~al.}(1999)\citenamefont {Keith},
  \citenamefont {Geimer}, \citenamefont {Jingwu},\ and\ \citenamefont
  {William}}]{keith1999optimization}%
  \BibitemOpen
  \bibfield  {author} {\bibinfo {author} {\bibfnamefont {D.~P.}\ \bibnamefont
  {Keith}}, \bibinfo {author} {\bibfnamefont {S.}~\bibnamefont {Geimer}},
  \bibinfo {author} {\bibfnamefont {T.}~\bibnamefont {Jingwu}},\ and\ \bibinfo
  {author} {\bibfnamefont {E.~B.}\ \bibnamefont {William}},\ }\bibfield
  {title} {\bibinfo {title} {Optimization of pelvic heating rate distributions
  with electromagnetic phased arrays},\ }\href@noop {} {\bibfield  {journal}
  {\bibinfo  {journal} {International Journal of Hyperthermia}\ }\textbf
  {\bibinfo {volume} {15}},\ \bibinfo {pages} {157} (\bibinfo {year}
  {1999})}\BibitemShut {NoStop}%
\bibitem [{\citenamefont {Lee}\ \emph {et~al.}(2023)\citenamefont {Lee},
  \citenamefont {Han}, \citenamefont {Song}, \citenamefont {Lee},\ and\
  \citenamefont {Kim}}]{lee2023discrete}%
  \BibitemOpen
  \bibfield  {author} {\bibinfo {author} {\bibfnamefont {S.}~\bibnamefont
  {Lee}}, \bibinfo {author} {\bibfnamefont {S.}~\bibnamefont {Han}}, \bibinfo
  {author} {\bibfnamefont {W.}~\bibnamefont {Song}}, \bibinfo {author}
  {\bibfnamefont {K.-J.}\ \bibnamefont {Lee}},\ and\ \bibinfo {author}
  {\bibfnamefont {S.}~\bibnamefont {Kim}},\ }\bibfield  {title} {\bibinfo
  {title} {Discrete source optimization for microwave hyperthermia using
  quantum annealing},\ }\href {https://doi.org/10.1109/LAWP.2023.3344792}
  {\bibfield  {journal} {\bibinfo  {journal} {IEEE Antennas and Wireless
  Propagation Letters}\ ,\ \bibinfo {pages} {1}} (\bibinfo {year}
  {2023})}\BibitemShut {NoStop}%
\bibitem [{\citenamefont {Boag}\ and\ \citenamefont
  {Leviatan}(1990)}]{boag1990optimal}%
  \BibitemOpen
  \bibfield  {author} {\bibinfo {author} {\bibfnamefont {A.}~\bibnamefont
  {Boag}}\ and\ \bibinfo {author} {\bibfnamefont {Y.}~\bibnamefont
  {Leviatan}},\ }\bibfield  {title} {\bibinfo {title} {Optimal excitation of
  multiapplicator systems for deep regional hyperthermia},\ }\href@noop {}
  {\bibfield  {journal} {\bibinfo  {journal} {IEEE transactions on biomedical
  engineering}\ }\textbf {\bibinfo {volume} {37}},\ \bibinfo {pages} {987}
  (\bibinfo {year} {1990})}\BibitemShut {NoStop}%
\bibitem [{\citenamefont {Boag}\ \emph {et~al.}(1993)\citenamefont {Boag},
  \citenamefont {Leviatan},\ and\ \citenamefont {Boag}}]{boag1993analysis}%
  \BibitemOpen
  \bibfield  {author} {\bibinfo {author} {\bibfnamefont {A.}~\bibnamefont
  {Boag}}, \bibinfo {author} {\bibfnamefont {Y.}~\bibnamefont {Leviatan}},\
  and\ \bibinfo {author} {\bibfnamefont {A.}~\bibnamefont {Boag}},\ }\bibfield
  {title} {\bibinfo {title} {Analysis and optimization of waveguide
  multiapplicator hyperthermia systems},\ }\href@noop {} {\bibfield  {journal}
  {\bibinfo  {journal} {IEEE transactions on biomedical engineering}\ }\textbf
  {\bibinfo {volume} {40}},\ \bibinfo {pages} {946} (\bibinfo {year}
  {1993})}\BibitemShut {NoStop}%
\bibitem [{\citenamefont {Ling}\ \emph {et~al.}(1984)\citenamefont {Ling},
  \citenamefont {Lee},\ and\ \citenamefont {Gee}}]{ling1984frequency}%
  \BibitemOpen
  \bibfield  {author} {\bibinfo {author} {\bibfnamefont {H.}~\bibnamefont
  {Ling}}, \bibinfo {author} {\bibfnamefont {S.-W.}\ \bibnamefont {Lee}},\ and\
  \bibinfo {author} {\bibfnamefont {W.}~\bibnamefont {Gee}},\ }\bibfield
  {title} {\bibinfo {title} {Frequency optimization of focused microwave
  hyperthermia applicators},\ }\href@noop {} {\bibfield  {journal} {\bibinfo
  {journal} {Proceedings of the IEEE}\ }\textbf {\bibinfo {volume} {72}},\
  \bibinfo {pages} {224} (\bibinfo {year} {1984})}\BibitemShut {NoStop}%
\bibitem [{\citenamefont {Vander~Vorst}\ \emph {et~al.}(2006)\citenamefont
  {Vander~Vorst}, \citenamefont {Rosen},\ and\ \citenamefont
  {Kotsuka}}]{vander2006rf}%
  \BibitemOpen
  \bibfield  {author} {\bibinfo {author} {\bibfnamefont {A.}~\bibnamefont
  {Vander~Vorst}}, \bibinfo {author} {\bibfnamefont {A.}~\bibnamefont
  {Rosen}},\ and\ \bibinfo {author} {\bibfnamefont {Y.}~\bibnamefont
  {Kotsuka}},\ }\href@noop {} {\emph {\bibinfo {title} {RF/microwave
  interaction with biological tissues}}}\ (\bibinfo  {publisher} {John Wiley \&
  Sons},\ \bibinfo {year} {2006})\BibitemShut {NoStop}%
\bibitem [{\citenamefont {Chew}(1995)}]{chew1995waves}%
  \BibitemOpen
  \bibfield  {author} {\bibinfo {author} {\bibfnamefont {W.~C.}\ \bibnamefont
  {Chew}},\ }\href@noop {} {\emph {\bibinfo {title} {Waves and fields in
  inhomogeneous media}}},\ Vol.~\bibinfo {volume} {16}\ (\bibinfo  {publisher}
  {IEEE Press},\ \bibinfo {year} {1995})\BibitemShut {NoStop}%
\bibitem [{\citenamefont {Harrington}(2001)}]{harrington2001time}%
  \BibitemOpen
  \bibfield  {author} {\bibinfo {author} {\bibfnamefont {R.~F.}\ \bibnamefont
  {Harrington}},\ }\href@noop {} {\emph {\bibinfo {title} {Time-harmonic
  electromagnetic fields}}}\ (\bibinfo  {publisher} {IEEE Press},\ \bibinfo
  {year} {2001})\BibitemShut {NoStop}%
\bibitem [{\citenamefont {Gabriel}\ \emph {et~al.}(1996)\citenamefont
  {Gabriel}, \citenamefont {Lau},\ and\ \citenamefont
  {Gabriel}}]{gabriel1996dielectric}%
  \BibitemOpen
  \bibfield  {author} {\bibinfo {author} {\bibfnamefont {S.}~\bibnamefont
  {Gabriel}}, \bibinfo {author} {\bibfnamefont {R.}~\bibnamefont {Lau}},\ and\
  \bibinfo {author} {\bibfnamefont {C.}~\bibnamefont {Gabriel}},\ }\bibfield
  {title} {\bibinfo {title} {The dielectric properties of biological tissues:
  Iii. parametric models for the dielectric spectrum of tissues},\ }\href@noop
  {} {\bibfield  {journal} {\bibinfo  {journal} {Physics in medicine \&
  biology}\ }\textbf {\bibinfo {volume} {41}},\ \bibinfo {pages} {2271}
  (\bibinfo {year} {1996})}\BibitemShut {NoStop}%
\bibitem [{\citenamefont {Iero}\ \emph {et~al.}(2015)\citenamefont {Iero},
  \citenamefont {Crocco},\ and\ \citenamefont {Isernia}}]{iero2015role}%
  \BibitemOpen
  \bibfield  {author} {\bibinfo {author} {\bibfnamefont {D.~A.}\ \bibnamefont
  {Iero}}, \bibinfo {author} {\bibfnamefont {L.}~\bibnamefont {Crocco}},\ and\
  \bibinfo {author} {\bibfnamefont {T.}~\bibnamefont {Isernia}},\ }\bibfield
  {title} {\bibinfo {title} {On the role and choice of source polarization in
  time-reversal focusing of vector fields},\ }\href@noop {} {\bibfield
  {journal} {\bibinfo  {journal} {IEEE Antennas and Wireless Propagation
  Letters}\ }\textbf {\bibinfo {volume} {15}},\ \bibinfo {pages} {214}
  (\bibinfo {year} {2015})}\BibitemShut {NoStop}%
\bibitem [{\citenamefont {Song}\ \emph {et~al.}(2022)\citenamefont {Song},
  \citenamefont {Sun}, \citenamefont {Du}, \citenamefont {Wu}, \citenamefont
  {Niu}, \citenamefont {Fu}, \citenamefont {Tan}, \citenamefont {Ren},
  \citenamefont {Chen},\ and\ \citenamefont {Meng}}]{song2022micro}%
  \BibitemOpen
  \bibfield  {author} {\bibinfo {author} {\bibfnamefont {J.}~\bibnamefont
  {Song}}, \bibinfo {author} {\bibfnamefont {X.}~\bibnamefont {Sun}}, \bibinfo
  {author} {\bibfnamefont {Y.}~\bibnamefont {Du}}, \bibinfo {author}
  {\bibfnamefont {Q.}~\bibnamefont {Wu}}, \bibinfo {author} {\bibfnamefont
  {M.}~\bibnamefont {Niu}}, \bibinfo {author} {\bibfnamefont {C.}~\bibnamefont
  {Fu}}, \bibinfo {author} {\bibfnamefont {L.}~\bibnamefont {Tan}}, \bibinfo
  {author} {\bibfnamefont {X.}~\bibnamefont {Ren}}, \bibinfo {author}
  {\bibfnamefont {L.}~\bibnamefont {Chen}},\ and\ \bibinfo {author}
  {\bibfnamefont {X.}~\bibnamefont {Meng}},\ }\bibfield  {title} {\bibinfo
  {title} {Micro-opening ridged waveguide tumor hyperthermia antenna combined
  with microwave-sensitive mof material for tumor microwave hyperthermia
  therapy},\ }\href@noop {} {\bibfield  {journal} {\bibinfo  {journal} {ACS
  Applied Bio Materials}\ }\textbf {\bibinfo {volume} {5}},\ \bibinfo {pages}
  {4154} (\bibinfo {year} {2022})}\BibitemShut {NoStop}%
\bibitem [{\citenamefont {Poon}\ \emph {et~al.}(2010)\citenamefont {Poon},
  \citenamefont {O'Driscoll},\ and\ \citenamefont {Meng}}]{meng2010optimal}%
  \BibitemOpen
  \bibfield  {author} {\bibinfo {author} {\bibfnamefont {A.~S.~Y.}\
  \bibnamefont {Poon}}, \bibinfo {author} {\bibfnamefont {S.}~\bibnamefont
  {O'Driscoll}},\ and\ \bibinfo {author} {\bibfnamefont {T.~H.}\ \bibnamefont
  {Meng}},\ }\bibfield  {title} {\bibinfo {title} {Optimal frequency for
  wireless power transmission into dispersive tissue},\ }\href
  {https://doi.org/10.1109/TAP.2010.2044310} {\bibfield  {journal} {\bibinfo
  {journal} {IEEE Transactions on Antennas and Propagation}\ }\textbf {\bibinfo
  {volume} {58}},\ \bibinfo {pages} {1739} (\bibinfo {year}
  {2010})}\BibitemShut {NoStop}%
\bibitem [{\citenamefont {Kim}\ \emph {et~al.}(2012{\natexlab{b}})\citenamefont
  {Kim}, \citenamefont {Ho}, \citenamefont {Chen},\ and\ \citenamefont
  {Poon}}]{kim2012wireless2}%
  \BibitemOpen
  \bibfield  {author} {\bibinfo {author} {\bibfnamefont {S.}~\bibnamefont
  {Kim}}, \bibinfo {author} {\bibfnamefont {J.~S.}\ \bibnamefont {Ho}},
  \bibinfo {author} {\bibfnamefont {L.~Y.}\ \bibnamefont {Chen}},\ and\
  \bibinfo {author} {\bibfnamefont {A.~S.}\ \bibnamefont {Poon}},\ }\bibfield
  {title} {\bibinfo {title} {Wireless power transfer to a cardiac implant},\
  }\href@noop {} {\bibfield  {journal} {\bibinfo  {journal} {Applied Physics
  Letters}\ }\textbf {\bibinfo {volume} {101}} (\bibinfo {year}
  {2012}{\natexlab{b}})}\BibitemShut {NoStop}%
\end{thebibliography}%
\end{document}